\documentclass[aps,twocolumn]{revtex4-1}

\usepackage{newlfont}
\usepackage{color}           
\usepackage{soul}            
\usepackage{amssymb}
\usepackage{amsfonts}
\usepackage{amsmath}
\usepackage{wasysym}
\usepackage{graphicx}
\usepackage{epsfig}
\usepackage{amsthm}
\usepackage{bm}
\usepackage[colorlinks=true, citecolor=blue, urlcolor=blue ]{hyperref}

\begin{document}


\title{Multisite Entanglement acts as a Better Indicator of Quantum Phase Transitions\\in Spin Models with Three-spin Interactions}

\author{Manabendra N. Bera, R. Prabhu, Aditi Sen(De) and Ujjwal Sen}
\affiliation{Harish-Chandra Research Institute, Chhatnag Road, Jhunsi, Allahabad 211 019, India}

\begin{abstract}

We employ a genuine multipartite entanglement measure, the generalized geometric measure, for investigating the  quantum phase transition 
in an  infinite quantum spin-1/2 chain with two-spin as well as three-spin interactions. We show that in contrast to bipartite entanglement of the ground state, the genuine multiparty one 
unerringly 
indicates the quantum phase transition in the system. 
The system lends itself to a complementary behavior between bipartite and genuine multipartite entanglements. 
We also investigate the trend of the multipartite entanglement in finite chains, in which we perform a finite-size scaling to confirm the 
signature of the quantum phase transition already in finite systems. The finite-size investigation renders the phenomenon to be observable in real systems.



\end{abstract}

\pacs{}

\maketitle

\section{Introduction}
Quantum entanglement is one of the basic ingredients in quantum physics and is identified as the key resource for many potential 
applications in quantum communication and computational tasks \cite{Horodecki09}. In recent years,
great
efforts have been put 
towards characterizing and quantifying the entanglement properties in quantum systems. However, a relatively good understanding  is limited 
to bipartite systems, with a general characterization and quantification of entanglement in multiparty  systems being still an open problem. 

On the other hand, it has been established in the last decade or so that there are connections between  bipartite entanglement \cite{Horodecki09} and quantum critical 
phenomena \cite{Subirbook11} in many-body systems. In particular,
it has been shown that bipartite entanglement can be useful for detecting (zero temperature) quantum phase transitions, which are transitions driven solely 
by quantum fluctuations between qualitatively distinct phases of quantum many-body systems \cite{Sen07,Amico08}.
The  bipartite entanglement 
measures, employed for this purpose, include concurrence \cite{Wootters98} and logarithmic negativity \cite{Werner02}.

With the vast developments in the field of simulating quantum many-body systems in different physical substrates, including 
atoms in an optical lattice \cite{Sen07,Bloch08}, trapped ions \cite{iontraprmp}, photons \cite{Ma11}, and nuclear magnetic resonance \cite{Peng09}, 
%
%
it is now possible to manipulate
and measure single party, bipartite, and multipartite properties of many-body quantum systems in the laboratory. 
This is performed by using the unprecedented levels of control that has been recently acquired in preparation and manipulation of 
such systems. 
%
Most of these investigations are restricted to Hamiltonians with two-body interactions.

Quantum many-body systems, however possess, in general, also three- and higher-body interactions, along with the two-body ones. 
Three-body interactions give rise to a whole new world of interesting physics 
\cite{Pachos05, Stoll05}. 
It has been shown that the multi-body interaction can lead to novel quantum phases with striking features 
\cite{Sandvik02}. Such many-body interactions have been observed, for example, in complex oxides
\cite{Monkman12}, as well as in  
engineered physical systems in the laboratory \cite{Delves59, Buchler07, Williams09, Lesanovsky11}. 
In particular, it was proposed that dominant three-body interactions can be created in a system of cold polar molecules \cite{Buchler07}.
Generation of effective three-body interactions was also shown to be possible by using the two-body collisions of atoms of a three-dimensional optical lattice \cite{Williams09}. 
Three-body interactions can potentially be simulated in one-dimensional atomic lattice gases with Rydberg atoms \cite{Lesanovsky11}.
Recently, Hamiltonians with multi-body interactions in ultracold atoms have been realized \cite{Mark11}.

In this paper, we consider the infinite spin-1/2  chain with 
%
 three-body interactions along with isotropic (two-body) $XY$ couplings. 
The entanglement properties of this model have been investigated in Refs. \cite{Yang05, Lou06}. 
In particular, it has been observed that the (bipartite) entanglement of nearest-neighbor spins cannot reliably indicate the critical behavior of this model \cite{Yang05}. We show that a 
genuine multiparty entanglement measure of the ground state can be used to faithfully indicate the quantum phase transition (QPT) 
present. The genuine multiparty entanglement employed is the generalized geometric measure (GGM) \cite{Aditi10}. 
 The calculations for the infinite chain are carried out analytically, by applying the Jordan-Wigner transformation on the spin Hamiltonian. 
Keeping in mind that an experimental verification of the presented results and effects can potentially be carried out in finite spin systems, we have investigated the status of the effects, observed for the infinite chain, in finite-sized chains also. 
A finite-size 
scaling analysis reveals that the effect observed in the infinite chain can also be reproduced in finite chains.

The paper is organized as follows. We first briefly introduce the genuine multipartite entanglement, the GGM, in Sec. \ref{sec:GGM}. In Sec. \ref{sec:ModTheo}, 
we present the Hamiltonian of the spin-1/2 $XX$ (i.e., isotropic \(XY\)) chain with three-spin interactions, that we are investigating. Sec. \ref{sec:ResDisc} presents the results 
obtained for the infinite and finite spin chains. Finally, we give a conclusion in Sec. \ref{sec:Conclusion}.

\section{\label{sec:GGM} Generalized geometric measure}
In this section, we introduce the genuine multiparticle entanglement measure which we shall be using to investigate the multipartite entanglement of 
the system under consideration. A multiparty pure quantum state is said to be \textit{genuinely} multiparty entangled
if it is entangled across every bipartition of its constituent parties. To quantify genuine multipartite entanglement present in a multiparty
system, we will use the generalized geometric measure \cite{Aditi10}. For an N-party pure quantum state $| \psi_N \rangle$, 
the GGM is defined as
\begin{equation}
 G(| \psi_N \rangle)=1-\max |\langle \phi_N|\psi_N\rangle|^2
\end{equation} 
where the maximization is carried out over all pure states $|\phi_N\rangle$ that are not genuinely N-party entangled. The above expression
can be simplified as 
\begin{equation}
 G(| \psi_N \rangle)=1-\max \{ \lambda^2\},
\label{eqn:GGM}
\end{equation} 
where $\{\lambda^2\}$ denotes the collection of the squares of all maximal Schmidt coefficients of all bipartite splits of the pure state $|\psi_N\rangle$.

\section{\label{sec:ModTheo}Model Hamiltonian}

The model Hamiltonian that we consider in this paper has two-body as well as three-body interactions. The two-body part is the well-known \(XX\) interaction, and is interesting on its own \cite{Subirbook11}. 
 With the introduction of the additional three-body terms, the model exhibits interesting physics, including chiral phases \cite{Pachos05}.
 The governing Hamiltonian is given by
%
\cite{Lou04, Lou05}
\begin{equation}
\begin{split}
H= & -\frac{J}{4}\sum_{n=1}^N \Big[ \sigma_n^x \sigma_{n+1}^x + \sigma_n^y \sigma_{n+1}^y + \\
   &   \frac{\alpha}{2} \left( \sigma_{n-1}^x \sigma_{n}^z \sigma_{n+1}^y - \sigma_{n-1}^y \sigma_{n}^z \sigma_{n+1}^x \right) \Big]
\end{split}
\label{eqn:Hinf}
\end{equation}
where $J$ is the coupling constant for the two-body interactions while 
$\alpha J/2$ is that for the three-body ones, 
$N$ is the total number of spins arranged in an one-dimensional array, and $\sigma_n^a \ (a=x,y,z)$ are the Pauli spin matrices at the site $n$. 
 Note that $\alpha $ is a dimensionless quantity 
characterizing the strength of the three-spin interaction.
Here the periodic boundary condition is assumed i.e., $\sigma_{N+1}=\sigma_1$. 
This Hamiltonian can be diagonalized in two steps. To that aim, we first apply a Jordan-Wigner transformation \cite{LSM61} given by
\begin{equation*}
\sigma_n^x=\prod_{m=1}^{n-1}\left(1-2c_m^{\dag}c_m\right)\left(c_n^{\dag}+c_n\right), \\
\end{equation*}  
\begin{equation*}
\sigma_n^y=\frac{1}{i}\prod_{m=1}^{n-1} \left(c_n^{\dag}-c_n\right) \left(1-2c_m^{\dag}c_m\right), 
\end{equation*}
\begin{equation*}
\sigma_n^z= 2c_n^{\dag}c_n -1,
\end{equation*}
whence the Hamiltonian takes the form
\begin{equation}
 H=J\sum_{n=1}^N \left[-\frac{1}{2} \left(c_n^{\dag} c_{n+1} + \mbox{h.c.} \right) + \frac{\alpha}{4 i} \left( c_n^{\dag} c_{n+2} - \mbox{h.c.} \right) \right].
\end{equation} 
The  boundary condition employed here 
is
either periodic or anti-periodic, 
depending on whether the number of spinless fermions is even or odd, 
respectively. 
We now apply the Fourier transformation 
\begin{equation*}
 c_n^{\dag}=\frac{1}{\sqrt{N}}\sum_k \exp \ (-ikn)c_k^{\dag}, \ \ c_n=\frac{1}{\sqrt{N}}\sum_k \exp \ (ikn)c_k
\end{equation*} 
where for an odd (even) number of spinless fermions, $k=(2\pi/N)m$  $(k=(2\pi/N)(m+1/2))$ in the subspaces with $m=-(N-1)/2,\ -(N-1)/2+1,...,(N-1)/2$ 
$(m=-N/2,\ -N/2+1,...,N/2-1)$.
As a consequence, the Hamiltonian reduces to
\begin{equation}
 H=\sum_k \varepsilon(k)c_k^{\dag}c_k.
\end{equation} 

The resulting energy dispersion is given by
\begin{equation}
 \varepsilon(k)=-J\left[\cos \ (k) -\frac{\alpha}{2} \sin \ (2k)  \right].
\end{equation} 
The $\varepsilon(k)$ has the following features: In the regime where $\alpha\leqslant\alpha_c=1$, 
only two Fermi points at $k_F=\pm\pi/2$ are present. But in the regime where $\alpha>\alpha_c$, there emerges two additional Fermi points in the negative-energy region in the k-space. So, four
Fermi points appear with the two additional points at $k^1_F= \arcsin(1/ \alpha)$ and $k^2_F=\pi-\arcsin(1/ \alpha)$. In the thermodynamic limit ($N\rightarrow\infty$), the ground state 
of the system corresponds to the configuration where all the states with $\varepsilon(k)\leqslant0$ are populated and the $\varepsilon(k)>0$ are empty. So, when 
$\alpha \leqslant \alpha_c$, the configuration of the ground state is of the states with $|k|<k_F$. On the other hand, when $\alpha > \alpha_c$, the ground state is formed by the states
with $-k_F<k<k_F^1$ and $k_F<k<k_F^2$ as filled. The energy per site of the ground state is
\begin{equation}
 E_0/N=\frac{1}{2\pi}\int_{\{k\}}\varepsilon(k)dk,
\end{equation} 
where the integration region $\{k\}=[-k_F,k_F ]$ for $\alpha \leqslant 1$ and $\{k\}=[-k_F,k^1_F]\cup[k_F,k^2_F]$ for $\alpha>1$. After a simple calculation, it takes the form
 \begin{equation*}
  E_0/JN=\left\lbrace 
 \begin{split}
 & -\frac{1}{\pi} \ \ \ \ \ \ \ \ \ \ \ \ \ \ \  \ \ \ \alpha < 1, \\ 
 & -\frac{1}{2 \pi}\left( \frac{1}{\alpha}+\alpha \right) \  \ \ \alpha \geq 1.
 \end{split}
 \right.
 \end{equation*} 
The ground state energy has a constant value for $\alpha < 1$ while it increases with $\alpha$ for $\alpha \geq 1$, and the QPT occurs at 
$\alpha=\alpha_c=1$ \cite{Lou04}. The magnetization for the ground state is zero for all values of $\alpha$. 

To evaluate the entanglement properties, one needs to fully characterize the ground state of the system. Once the expectation values of the two-point correlation 
functions $\langle c_n^{\dag} c_m \rangle=f_{n,m}$ are known, the ground state of the model can be completely 
characterized, and any other expectation value can be evaluated by using the Wick's theorem \cite{LSM61}. The $f_{n,m}$'s for an infinite chain can be calculated to
form a correlation matrix \(F\), which characterizes the density matrix of the ground state \cite{Lou06}: 
\begin{equation}
f_{n,m}=\langle c_n^{\dag} c_m \rangle=\frac{1}{2\pi}\int_{\{k\}} \exp \left[ -ik(m-n)\right] dk, 
\end{equation} 
where the integration is again carried out in the region $\{k\}=[-k_F,k_F ]$ for $\alpha \leqslant 1$ and $\{k\}=[-k_F,k^1_F]\cup[k_F,k^2_F]$ for $\alpha>1$.
The analytical expression of the two-point correlation function  is given by
\begin{equation}
 f_{n,n+m}= \left\lbrace 
\begin{split}
& \frac{1}{m \pi} \sin\left(\frac{m \pi}{2}\right) \ \ \ \ \ \ \ \ \ \ \ \ \ \ \ \ \ \ \ \ \ \ \ \ \ \ \ \ \ \ \ \alpha < 1 , \\
& \frac{1}{2 m \pi} [1-(-1)^m] \sin \left[m \arcsin \left(\frac{1}{\alpha}\right) \right]\ \alpha \geqslant 1 ,
\end{split}
\right.
\label{eqn:fij}
\end{equation} 
for $m\neq 0$ and 
\begin{equation}
 f_{n,n}=\frac{1}{2}.
\label{eqn:f11}
\end{equation} 
When $m$ is an even integer, the $f_{n,n+m}$ always vanishes.
The elimination of the rows and columns from the \(F\)-matrix corresponds to the spins which do not belong to the block, and
hence the \(n\)-particle reduced density matrices (which we denote as $\rho_n$) of the ground state can be characterized by using the reduced correlation matrix 
\begin{equation}
 F_n=
\begin{pmatrix}
 f_{1,1} & f_{1,2} & \cdots & f_{1,n} \\
  f_{2,1} & f_{2,2} & \cdots & f_{2,n} \\
  \vdots  & \vdots  & \ddots & \vdots  \\
  f_{n,1} & f_{n,2} & \cdots & f_{n,n}
\end{pmatrix}.
\label{eqn:Fn}
\end{equation} 

The resulting reduced density matrices, in terms of the spinless fermion operators, takes the form \cite{Its08}
\begin{equation}
 \rho_n=\prod_{n=0}^n\left( \lambda_n c_n^{\dag}c_n + (1-\lambda_n) c_nc_n^{\dag} \right)
\end{equation} 
where $\lambda_n$ is the $n^{\mathrm th}$ eigenvalue of the correlation matrix $F_n$ and the values $\lambda_n$ and $1-\lambda_n$ are the eigenvalues of the 
reduced density matrices.




\section{\label{sec:ResDisc}Results: GGM vs bipartite entanglement}

We now demonstrate the efficacy of the generalized geometric measure in detecting the quantum phase transition in the three-body interaction Hamiltonian (Eq. (\ref{eqn:Hinf})). We begin by considering the infinite chain. Later, we will also consider the 
situation for finite ones.  

\subsection{Infinite spin chain}

The two-site reduced density matrix $\rho_{m,m+1}$ (after taking the partial trace over all other sites except $m$ and $m+1$) of the ground state, can be expressed as
\begin{equation}
 \rho_{m,m+1}=
\begin{pmatrix}
 u & 0 & 0 & 0 \\
  0 & z & y & 0 \\
  0  & y  & z & 0  \\
  0 & 0 & 0 & u
\end{pmatrix},
\end{equation} 
where the elements can be calculated from the correlation functions:
\begin{equation}
\begin{split}
 u &=\frac{1}{4}\left( 1 +\langle \sigma^z_m \sigma^z_{m+1} \rangle \right)\\
 &=\left(\frac{1}{2} +f_{1,2} \right) \left(\frac{1}{2} -f_{1,2} \right),\\
z & =\frac{1}{4}\left( 1- \langle \sigma^z_m \sigma^z_{m+1} \rangle \right)\\
& = \frac{1}{4}+(f_{1,2})^2,\\
y &=\frac{1}{4}\left( \langle \sigma^x_m \sigma^x_{m+1} \rangle + \langle \sigma^y_m \sigma^y_{m+1} \rangle \right)\\
& = f_{1,2},
\end{split}
\end{equation} 
where the averages are carried out in the ground state.

\begin{figure}
\centering 
\includegraphics[width=0.38\textwidth, angle=0]{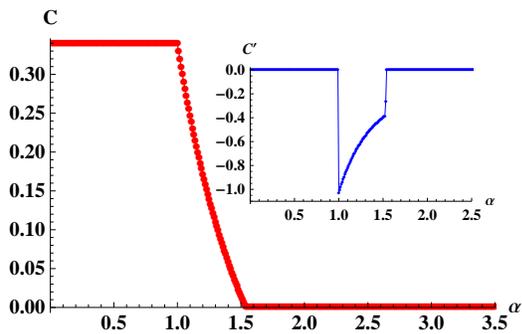}
\caption{\label{fig:ConInf} (Color online.) The failure of bipartite entanglement to indicate QPT in spin Hamiltonians with three-body interactions.
 The figure represents the variation of concurrence ($C$) with respect to $\alpha$. The concurrence (in ebits) is represented on the vertical axis and \(\alpha\) (dimensionless) is represented on the horizontal axis. 
The sharp change of $C$ at $\alpha=1$  corresponds to a QPT. There is  a second sharp change 
at $\alpha=\pi/2$, which however does  not correspond to any QPT. 
The inset figure depicts the variation of the first derivative ($C'$) of concurrence with respect to $\alpha$.
}
\end{figure}

We can now calculate the bipartite entanglement of the two-party reduced state, obtained from the ground state,
 by using entanglement measures like concurrence.
The concurrence of a two-qubit state \(\rho_2\) is defined as
\begin{equation}
 C= \max \{ 0,\lambda_1-\lambda_2-\lambda_3-\lambda_4 \},
\end{equation} 
where the $\lambda_i$'s are the eigenvalues, with \(\lambda_1\) being the largest, of the Hermitian matrix $R=\sqrt{\sqrt{\rho_2}\widetilde{\rho_2}\sqrt{\rho_2}}$. 
Here, $\widetilde{\rho_2}=(\sigma_y \otimes \sigma_y) \rho^\ast_2 (\sigma_y \otimes \sigma_y)$, where the asterisk indicates a complex conjugation in the computational (i.e. \(\sigma^z\)) basis.
The behavior of concurrence of the two-particle reduced density matrix of the ground state with respect to $\alpha$ is depicted in Fig. \ref{fig:ConInf}. The derivative of concurrence 
with respect to $\alpha$ shows  discontinuities (see inset of Fig. \ref{fig:ConInf})  at 
$\alpha=1$ and $\alpha=\pi/2$, although the phase transition in the model occurs at $\alpha=1$. Hence it is clear that 
in this three-body interaction Hamiltonian, bipartite entanglement cannot faithfully capture  the quantum cooperative properties of the many-body system.
 More precisely, it signals to
a QPT which is not present in the quantum system \cite{Yang05}.

Cooperative phenomena like quantum phase transitions in a many-body system occurs due to the interplay among a large number of (if not, all the) particles of the system. This is despite the fact that the couplings of 
a given particle of the system 
are typically (although, not always) with a few ``nearby'' particles. The effect is arguably like in entanglement swapping, where two particles can get entangled even if they have never interacted in the past, but have interacted with two other 
particles respectively who in turn have interacted in the past \cite{Zukowski93}. Since a large number of particles are involved in creating the quantum phase transition, it is 
natural to expect that it will be signaled by a multisite physical property of the system, like a genuine multipartite entanglement measure. 
%

Towards this aim, let us now consider the behavior of the GGM, which is a genuine multipartite entanglement measure, in the ground state of the three-body interaction Hamiltonian given in Eq. (\ref{eqn:Hinf}).
As we have discussed in Sec. \ref{sec:GGM}, to evaluate the GGM, one has to
calculate the Schmidt coefficients of different bipartitions. Before calculating GGM, for ease of calculation, let us introduce here a multiparty entanglement measure,
 based on the GGM, for an \(N\)-party pure quantum state \(|\psi_N\rangle\), as
\begin{equation}
 G_n(|\psi_N\rangle)=1-\max\{\lambda_{n:\mathrm{\scriptsize{rest}}}^2\}
\end{equation} 
where $\{\lambda_{n:\mathrm{\scriptsize{rest}}}^2\}$ denotes the set containing  the squares of the Schmidt coefficients of the state $|\psi_N\rangle$ in a partition of the \(N\) constituent parties into a bipartition of \(n\) and \(N-n\) parties. 
This is still a multiparty entanglement measure (which we refer to as the ``\(n\)-th order geometric measure''), 
and can be used to reveal information about the GGM of the \(N\)-party state. The latter holds due to the fact that the minimum of $G_n$'s corresponds 
to the GGM of the multiparty state of \(N\) parties. 
Since the interactions in the Hamiltonian are between nearest two spins or between nearest three spins, it is plausible that the most interesting physics will be seen in the \(G_n\) obtained from nearest \(n\) spins, and the 
same is referred to as \(G_n\) below.
With the help of the Eqs. (\ref{eqn:fij}), (\ref{eqn:f11}), and  (\ref{eqn:Fn}), we are able to find the analytical form 
for different $G_n$'s in the bipartitions.
In the case of bipartition of 2 spins versus the remaining spins (``2:rest''), we have
\begin{equation}
 G_2= \left\lbrace 
\begin{split}
& \frac{1}{2}-\frac{1}{\pi} \ \ \ \ \ \alpha < 1, \\
& \frac{1}{2}-\frac{1}{\pi \alpha} \ \ \ \alpha \geqslant 1.
\end{split}
\right.
\label{eqn:G2}
\end{equation} 
Similarly for \(n=3\) and \(n=4\), we respectively have 
\begin{equation}
 G_3= \left\lbrace 
\begin{split}
& \frac{1}{2}-\frac{\sqrt{2}}{\pi} \ \ \ \ \alpha < 1, \\
& \frac{1}{2} - \frac{\sqrt{2}}{\pi \alpha} \ \ \ \ \alpha \geqslant 1,
\end{split}
\right.
\label{eqn:G3}
\end{equation} 
and 
\begin{equation}
 G_4= \left\lbrace 
\begin{split}
& \frac{1}{2}-\frac{1+\sqrt{13}}{3 \pi} \ \ \ \ \ \ \ \ \ \ \ \ \ \ \ \ \ \ \alpha < 1, \\
& \frac{1}{2}-\frac{1}{\pi \alpha} + \frac{2 - \sqrt{4+9 \alpha^4}}{3 \pi \alpha^3}  \ \ \ \alpha \geqslant 1.
\end{split}
\right.
\label{eqn:G4}
\end{equation}

\begin{figure}
\centering 
\includegraphics[width=0.4\textwidth, angle=0]{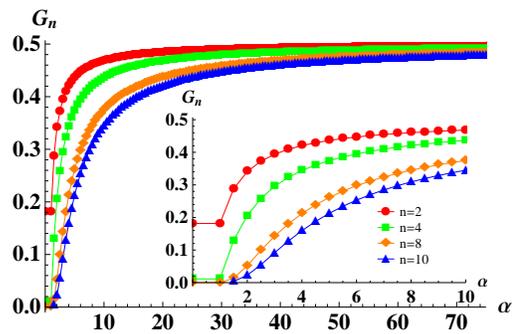}
\caption{\label{fig:GGMvsAlpha} (Color online.) Generalized geometric measure detects quantum phase transition. The figure represents the trend of the \(n\)-th order geometric measures ($G_n$), on the vertical axis,  
 with respect to $\alpha$ on the horizontal axis.  The inset depicts the same figure, but by zooming in near the phase transition. 
The \(G_n\)'s increases monotonically after the transition and reaches \(1/2\) for large  $\alpha\). 
All quantities are dimensionless.
}

\end{figure}


It is clear from the expressions of $G_2, \ G_3$, and $G_4$ that whatever bipartition we consider, the minimum value of the set $\{ G_2, \ G_3, \ ... \}$, 
which correspond to the GGM of the ground state, is a constant for $\alpha<\alpha_c=1$, and for  $\alpha \geq \alpha_c$, it always increases as can be seen from 
Fig. \ref{fig:GGMvsAlpha}. 
Note that, the $G_n$'s have the order $G_2>G_3>G_4 >\hdots $, and they merge in the 
asymptotic limit $\alpha\rightarrow\infty$, to reach 
the value $1/2$. 

In the regime $\alpha<\alpha_c=1$, the GGM can be expressed in the form
\begin{equation}
 G_{n}=1/2-A_n,
\end{equation} 
where the $A_n$ is a constant (with respect to \(\alpha\)) which only depends on the $n$. The constant $A_n$ slowly increases with the increase in $n$ (see Table I).
\begin{table}%
\begin{tabular}{|c|c|}
\hline
$n$ & $A_n$ \\
\hline
2 & 0.318309 \\
3 & 0.450158 \\
4 & 0.488664 \\
5 & 0.497669 \\
6 & 0.499544 \\
7 & 0.499913 \\
$n\rightarrow \infty$ & 0.5\\
\hline
\end{tabular}
\caption{The behavior of \(A_n\) with respect to \(n\).}
\label{}
\end{table}
Hence it clearly converges to $1/2$ as $ n \rightarrow \infty$. Therefore for  $\alpha<\alpha_c=1$, 
$G_{n}\rightarrow 0$ as $ n \rightarrow \infty$. \textit{So, for $\alpha<1$, genuine multipartite entanglement is absent while a finite bipartite entanglement is present in the system.}

Let us now consider the case when $\alpha \geqslant \alpha_c=1$. In this regime, the GGM ($G_{n}$) depends on $\alpha$, and
can be written in the form 
\begin{equation}
 G_{n}=1/2-O(1/\alpha).
\end{equation} 
We find, from the expression of $F_n$ defined in Eq. (\ref{eqn:Fn}), that as $ \alpha \rightarrow \infty$, 
$G_{n}\rightarrow 1/2$ irrespective of $n$. 
\textit{So, when $ \alpha \rightarrow \infty$, the system becomes fully multiparty entangled with a vanishing bipartite entanglement.}

\begin{figure}
\centering 
\includegraphics[width=0.4\textwidth, angle=0]{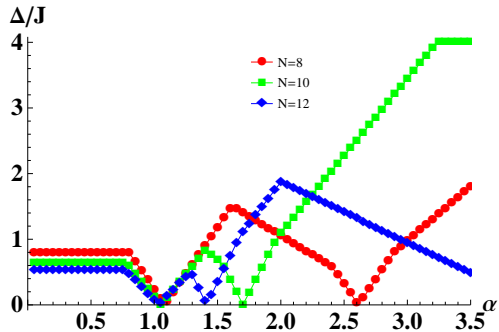}
\caption{\label{fig:DelEFin} (Color online.) Energy gaps for finite systems. We plot the energy gaps, \(\Delta/J\), for different finite-size spin chains against the system parameter \(\alpha\).
Both axes represent dimensionless quantities.
}
\end{figure}


\begin{figure}
\centering 
\includegraphics[width=0.4\textwidth,height=0.18\textheight, angle=0]{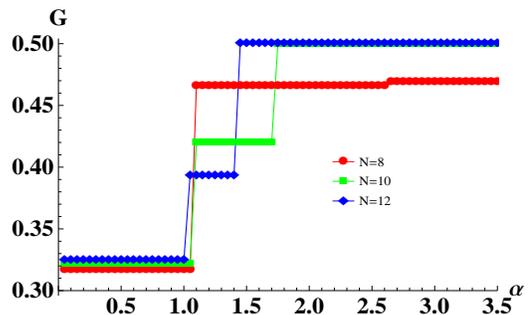}
\caption{\label{fig:GgmFin} (Color online.) The GGM for finite-size systems. 
We plot the GGM for systems of different finite spins (\(=N\)) on the vertical axis, against the system parameter \(\alpha\) on the horizontal one. 
Both quantities are dimensionless. 
See text for the finite-size scaling analysis for the transitions.
}
\end{figure}

\subsection{Finite spin chains}

Recent experimental developments strongly indicate that the Hamiltonians considered in this paper can be experimentally observed and the behavior of entanglement can be studied 
in a real system. For example, it has already been proposed that by using two-component ultracold Bose-Bose and Fermi-Fermi gas mixture and by suitably tuning 
the scattering length, the $XX$ Hamiltonian can be realized in ultracold atoms \cite{Sen07, Bloch08}. Moreover the additional
three-body interaction has also been created in the laboratories \cite{Mark11}. 

Keeping such possibilities in mind, in this section, we study 
the behavior of genuine multipartite entanglement in controllable finite spin systems, which are achievable by using current technologies. 
The Hamiltonian that we consider here is same as in Eq. (\ref{eqn:Hinf}). 
We perform exact diagonalization  to explore the properties of the ground state of the system. 
The spin chains with an odd number of spins have doubly degenerate ground states, while that is not the case for one with an even number of spins.
The degeneracy in the case of an odd number of spins can be removed with the introduction of a small transverse field. 


The ground states have a vanishing total magnetization for the whole range of $\alpha$. 
The energy difference ($\Delta$) between the 
ground state and the first excited state  has been calculated and the variation is shown in Fig. \ref{fig:DelEFin}. In the thermodynamic limit, a quantum phase 
transition in general occurs with the vanishing of the energy gap.
However, finite size simulations in our case show that there are several values of $\alpha$ where the $\Delta \rightarrow 0$. 
It turns out that the derivatives of genuine multipartite entanglement measure (GGM) also show discontinuities at the points where 
\(\Delta \rightarrow 0\), i.e. at points additional to the one near $\alpha=1$ (see Fig. \ref{fig:DelEFin}).

The derivative of the GGM exhibits a discontinuity near $\alpha=\alpha_c^N$
with \(\alpha_c^N\) approaching to $1$ 
with an increase in the number of spins. This is the signature of the quantum phase transition at \(\alpha = \alpha_c = 1\) in the infinite chain. 
By performing a finite-size scaling, we find that the point of discontinuity present in the finite system, approaches to $\alpha_c = 1$ as 
\begin{equation}
\alpha \sim \alpha_c + N^{-1.787}.                                                                                                                                                             
\end{equation}
Here \(N\) is the number of spins.
%
%
%
There appears  second discontinuities (for different $N$), seen in Fig. \ref{fig:GgmFin}, and these are away from $\alpha=1$ for relatively small \(N\). 
For example, for $N=8,\,10,\,\mbox{and}\,12$ 
the second ones occur at $\alpha=2.6, \ 1.7$, and $1.4$ respectively. It is clear that the second discontinuity point approaches to $\alpha=1$ in the large system limit. 
Again a  finite-size scaling is employed, and we find that the point of the second discontinuities (\(\alpha_2^N\))  approaches to $\alpha_c(=1)$ as 
\begin{equation}
 \alpha_2^N \sim \alpha_c+N^{-3.4}.
\end{equation} 



\section{\label{sec:Conclusion}Conclusion}
We have investigated the advantage of using the genuine multiparty entanglement measure called  generalized geometric measure for detecting quantum phase transitions in 
the infinite quantum spin-1/2 chains with two-spin isotropic \(XY\) interactions and 
three-spin interactions.
We find that in contrast to bipartite entanglement, like concurrence, which signals to phase transitions that are not present, the generalized geometric measure faithfully
detects the quantum phase transition in the system. 
Recent experimental achievements that have made it possible to realize quantum spin models in a controlled way, motivated us to consider the behavior of
the genuine multiparty entanglement for finite spin chains. We reveal that the generalized geometric measure can successfully notice changes in 
the finite systems, by analyzing the 
scaling behavior of which, we can detect the quantum phase transition in the systems.

\begin{acknowledgments}
R.P. acknowledges support from the Department of Science and Technology, Government of India, in the form of an INSPIRE faculty 
scheme at the Harish-Chandra Research Institute (HRI), India. We acknowledge computations performed at the cluster computing facility in HRI.
\end{acknowledgments}




\end{document}